\documentclass[final,sort&compress]{aipproc}

\layoutstyle{6x9}

\usepackage{amsmath,amssymb}
\usepackage{latexsym} 

\begin{document}

\newcommand{\nn}{\nonumber}
\newcommand{\be}{\begin{equation}}
\newcommand{\ee}{\end{equation}}
\newcommand{\ba}{\begin{eqnarray}}
\newcommand{\ea}{\end{eqnarray}}
\newcommand{\la}{\langle}
\newcommand{\ra}{\rangle}
\title{TMDs and Azimuthal Spin Asymmetries in a Light-Cone Quark Model}

\classification{
      13.88.+e, 
      12.39.-x,13.85.Ni,13.60.-r}
\keywords      {transverse momentum dependent parton distributions, azimuthal spin asymmetries}

\author{B. Pasquini}
{
  address={
Dipartimento di Fisica Nucleare e Teorica, Universit\`{a} degli Studi di Pavia, and\\
Istituto Nazionale di Fisica Nucleare, Sezione di Pavia, I-27100 Pavia, Italy}
}

\author{S. Boffi}{
  address={
Dipartimento di Fisica Nucleare e Teorica, Universit\`{a} degli Studi di Pavia, and\\
Istituto Nazionale di Fisica Nucleare, Sezione di Pavia, I-27100 Pavia, Italy}
}

\author{A.~V.~Efremov}
{
  address={
Joint Institute for Nuclear Research, Dubna,141980 Russia}
}

\author{P.~Schweitzer}{
  address={Department of Physics, University of Connecticut, Storrs, CT 06269, USA}
}

\begin{abstract}
 The main properties of the leading-twist transverse momentum 
dependent parton distributions  in a light-cone constituent quark model 
of the nucleon are reviewed, with focus on the role of the spin-spin and 
 spin-orbit correlations of quarks.
Results for azimuthal single spin 
asymmetries in semi-inclusive deep inelastic scattering are also discussed.
\end{abstract}

\maketitle


\section{TMDs and the Light-Cone CQM}

In recent years much work has been devoted to study semi-inclusive deep 
inelastic scattering (SIDIS), Drell-Yan  dilepton production and hadron production in $e^+ e^-$ annihilation as 
powerful tools to understand the nucleon structure. According to the 
factorization theorem, 
the physical 
observables of such processes can be expressed as convolution of hard partonic 
scattering cross sections, parton distribution functions (PDFs) and/or
 fragmentation functions 
(FFs)~\cite{Collins:2003fm,Collins:1992kk,Mulders:1995dh,Boer:1997nt,Brodsky:2002cx,Collins:2002kn}.
With respect to the usual 
inclusive deep inelastic scattering (DIS) where PDFs only depend on the 
longitudinal momentum fraction carried by the parton, now PDFs, as well as FFs,
 also depend on the transverse momentum.
 At leading twist there are eight transverse momentum dependent 
PDFs (TMDs)~\cite{Mulders:1995dh,Boer:1997nt}, three of them 
surviving when integrated over the transverse momentum and giving rise to 
the familiar parton density, helicity and transversity distributions. 
Data~\cite{Arneodo:1986cf,Diefenthaler:2005gx,Airapetian:1999tv,Airapetian:2001eg,Avakian:2003pk,Airapetian:2002mf,Airapetian:2006rx,Airapetian:2004tw,Alexakhin:2005iw,Kotzinian:2007uv} on SIDIS are available and many more are expected to come in future, 
giving first insights on the TMDs
\cite{Efremov:2002ut,Efremov:2004tp,Vogelsang:2005cs,Efremov:2006qm,Anselmino:2007fs,Arnold:2008ap,Anselmino:2008sg,Zhang:2008nu}.
However, model calculations
\cite{Jakob:1997wg,Avakian:2008dz,Pasquini:2008ax,Bacchetta:2008af,Meissner:2007rx,Yuan:2003wk,Gamberg:2007gb,Schweitzer:2001sr,Pasquini:2005dk,Efremov:2004tz} play an important role 
for unraveling the information on the quark dynamics encoded
in these novel functions. In this contribution we will review the results for the TMDs in a light-cone constituent quark model (CQM) which was successfully applied also for the calculation of the electroweak properties of the nucleon~\cite{Pasquini:2007iz} and
generalized parton distributions~\cite{Boffi:2007yc}.
\newline\noindent
A convenient way to describe parton distributions is to 
use the representation in terms of overlaps of light-cone wave functions
(LCWFs), which are the probability amplitudes to find a given $N$-parton configuration in the Fock-space expansion of the hadron state. 
This representation becomes useful in 
phenomenological applications where one can reasonably truncate the
expansion of the hadron state to the Fock components with a few partons.
In our approach, we consider the minimum Fock sector with just three-valence quarks. This truncation allows to describe the parton distributions in those kinematical region where the valence degrees of freedom are effective, while
the contributions from sea quarks and gluons
are suppressed.
The three-quark component of the LCWF, keeping the full transverse momentum dependence of the partons, can be classified in a model independent way in terms
of six independent light-cone amplitudes~\cite{Ji:2002xn}, which serve to parametrize the 
contribution
from the four different orbital angular momentum components $L_z$ compatible
with total angular momentum conservation, i.e.  $L_z=0,\pm 1 , 2$.
In Ref.~\cite{Pasquini:2008ax}, these six amplitudes have been explicitly derived
in a light-cone CQM, 
considering the relativistic spin dynamics arising from the boost of 
instant-form wave function into the light-cone.
 The instant-form wave function is constructed as a product of a momentum 
wave function which is in a pure S-wave state and invariant under permutations,
 and a spin-isospin wave function determined by SU(6)
symmetry.
The corresponding solution in light-cone dynamics is obtained through the
unitary  Melosh rotations acting
on the spin of the individual quarks.
The relativistic effects of the Melosh rotations
are evident in the presence
of spin-flip terms generating non-zero orbital 
angular momentum components which fit the 
model-independent classification of the three-quark LCWF~\cite{Pasquini:2008ax,Ji:2002xn}.
The explicit expressions of these light-cone amplitudes can be found in 
Ref.~\cite{Pasquini:2008ax}, while the corresponding results for the time-even 
TMDs are
\begin{eqnarray}
\label{eq:f1}
f^a_1(x,p_T)&=&
N^a \int{\rm d}[X]\
   \delta(x-x_3)\delta({\bf p}_{T}-{\bf p}_{\perp\,3})\
\vert \psi(\{x_i,{\bf p}_{\perp\,i}\})\vert^2,
\nn\\
\label{eq:g1}
g^a_{1L}(x,p_T)&=&
P^a\int{\rm d}[X]\
   \delta(x-x_3)\delta({\bf p}_{T}-{\bf p}_{\perp\,3})\
\frac{(m+ x M_0)^2 -{\bf p}^2_{T}}{(m+ xM_0)^2 + {\bf p}^2_{T}}\;
\vert \psi(\{x_i,{\bf p}_{\perp\,i}\})\vert^2,
\nn\\
\label{eq:g1T}
g^{a}_{1T}(x,p_T)&=&
P^a
\int{\rm d}[X]\
   \delta(x-x_3)\delta({\bf p}_{T}-{\bf p}_{\perp\,3})\
\frac{2M(m+ xM_0)}{(m+ xM_0)^2 + {\bf p}^2_{T}}\;
\vert \psi(\{x_i,{\bf p}_{\perp\,i}\})\vert^2,
\nn\\
h^{\perp\, a}_{1L}(x,p_T)&=&
- P^a
\int{\rm d}[X]\
   \delta(x-x_3)\delta({\bf p}_{T}-{\bf p}_{\perp\,3})\
\frac{2M(m+ xM_0)}{(m+ xM_0)^2 + {\bf p}^2_T}\;
\vert \psi(\{x_i,{\bf p}_{\perp\,i}\})\vert^2,\nonumber
\label{eq:h1L}
\\
\label{eq:h1T}
h^{\perp\,a}_{1T}(x,p_T)&=&-
P^a
\int{\rm d}[X]\
   \delta(x-x_3)\delta({\bf p}_{T}-{\bf p}_{\perp\,3})\
\frac{2M^2}{(m+ xM_0)^2 + {\bf p}^2_{T}}\;
\vert \psi(\{x_i,{\bf p}_{\perp\,i}\})\vert^2,
\nn\\
h^a_1(x,p_T)&=&
P^a
\int{\rm d}[X]
   \delta(x-x_3)\delta({\bf p}_{T}-{\bf p}_{\perp\,3})
\frac{(m+ xM_0)^2}{(m+ xM_0)^2 + {\bf p}^2_{T}}
\vert \psi(\{x_i,{\bf p}_{\perp\,i}\})\vert^2,
\label{eq:h1}
\end{eqnarray}
where 
the integration measure is defined as in Ref.~\cite{Pasquini:2008ax},
$M_0$ is 
the mass of the non-interacting three-quark system, and
 the flavor dependence is given by 
the factors $N^u=2$, $N^d=1,$ and  $P^u=4/3$,  
$P^d=-1/3$, as dictated by SU(6) symmetry.
A further consequence of the assumed SU(6) symmetry is the factorization
in Eqs.~(\ref{eq:h1}) of the momentum-dependent wave function
$\psi(\{x_i,{\bf p}_{\perp\,i}\})$  
from the spin-dependent factor arising from the Melosh rotations.
As a result one finds the following relations
\begin{eqnarray}
\label{eq:61}
&&2h^a_1(x,p_T)
=g^a_{1L}(x,p_T)+\frac{P^a}{N^a}f^a_1(x,p_T),
\qquad
h_{1L}^{\perp a}(x,p_T)
=-g_{1T}^a(x,p_T),
\\
&&
\frac{P^a}{N^a}f^a_1(x,p_T)
=h_1^a(x,p_T) -\frac{p_T^2}{2M^2}h_{1T}^{\perp \,a}(x,p_T).
\label{eq:61a}
\end{eqnarray}
These relations are common to several quark model calculations 
~\cite{Avakian:2008dz,Pasquini:2008ax,Jakob:1997wg,Efremov:2004tz}
though not all~\cite{Bacchetta:2008af}.
The common feature of such models is that gluon degrees of freedom are 
neglected. On the other side,
the recent model calculation of Ref.~\cite{Efremov:2004tz} 
found the interesting result that SU(6) symmetry is not a necessary condition
for the relation ~(\ref{eq:61a}).


\begin{figure}[t]
  \includegraphics[width=12.1 truecm]{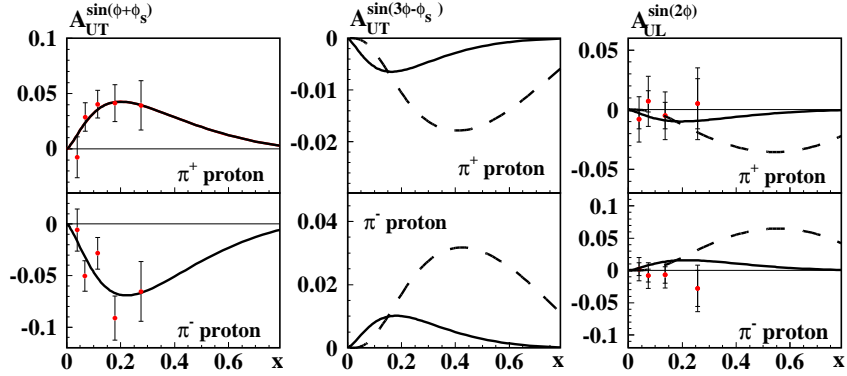}
  \caption{The SSAs $A_{UT}^{\sin(\phi+\phi_S)}$ (left column),
$A_{UT}^{\sin(3\phi-\phi_S)}$ (middle column), and
$A_{UL}^{\sin(2\phi)}$ (right column) in DIS production of charged pions
off proton target, as function of $x$. The solid curves show the results on the basis of the light-cone CQM evolved to the scale $Q^2=2.5 $ GeV$^2$, 
and the dashed
curves correspond to the predictions at the low scale of the model. 
The experimental data are from Refs.~\cite{Diefenthaler:2005gx,Airapetian:1999tv}.
\label{fig1}}
\end{figure}

\section{Results for azimuthal SSAs}
The results  in Eqs.~(\ref{eq:h1}) are general and can be applied 
to any CQM adopting the appropriate nucleon wave function.
In the following we will take the momentum wave-function 
from Schlumpf~\cite{Schlumpf:1992ce}.
In Fig.~\ref{fig1} we show the results for the single spin asymmetries
(SSAs) with unpolarized (U) beam
and transversely (T) polarized proton target in SIDIS of positive (upper panels) and negative (lower panels) pions.
The asymmetries $A_{UT}^{\sin(\phi+\phi_S)}$ (left column), 
$A_{UT}^{\sin(3\phi-\phi_S)}$ (middle column), 
and
$A_{UL}^{\sin(2\phi)}$ (right column) are due to the Collins function and to the three chirally-odd TMDs 
$h_1$,  $h_{1T}^\perp$, and $h_{1L}^\perp$, respectively. 
For the Collins function we use the results extracted in \cite{Efremov:2006qm}.
In the denominator of the asymmetries we take $f_1$ from~\cite{Gluck:1998xa}
and the unpolarized FF from~\cite{Kretzer:2000yf}, both valid at the scale $Q^2=2.5$ GeV$^2$.
The model results for $h_1$ evolved from the low hadronic scale
of the model to $Q^2=2.5 $ GeV$^2$ ideally describe the HERMES 
data~\cite{Diefenthaler:2005gx} for
$A_{UT}^{\sin(\phi+\phi_S)}$. 
This is in line with the favourable comparison 
between our model predictions~\cite{Pasquini:2005dk}  and the phenomenological extraction 
of the transversity and the tensor charges in Ref.~\cite{Anselmino:2007fs}.
In the case of $A_{UL}^{\sin(2\phi)}$ and $A_{UT}^{\sin(3\phi-\phi_S)}$ we compare the results obtained using the TMDs 
at the scale of the model (dashed curves) and the TMDs evolved 
at leading order to $Q^2=2.5$ GeV$^2$ (solid curves) 
using the evolution pattern of the transversity.
Although this is not the correct evolution pattern, 
it may give us a rough insight on the possible size of effects 
due to evolution 
(for a more detailed discussion we refer to~\cite{BEPS09}).
In the case of  $A_{UT}^{\sin(3\phi-\phi_S)}$, the evolution effects give 
smaller  asymmetries in absolute value and shift the peak at lower $x$ values.
Measurements in  range $0.1\lesssim x \lesssim 0.6$ are planned with the CLAS 12 GeV upgrade~\cite{Avakian-LOI-CLAS12} 
and will be able to discriminate between the two scenarios.
In the  region $x\lesssim 0.2$, there exist also preliminary 
deuteron target data~\cite{Kotzinian:2007uv}
which are compatible, within error bars,
with the model predictions both at the hadronic and the evolved scale.
Similar conclusions can be drawn also in the case of $A_{UL}^{\sin(2\phi)},$
 where we compare our results with HERMES data~\cite{Airapetian:1999tv}.



\bibliographystyle{aipproc}   

\bibliography{sample}


\bibliographystyle{aipproc}   

\IfFileExists{\jobname.bbl}{}
 {\typeout{}
  \typeout{******************************************}
  \typeout{** Please run "bibtex \jobname" to optain}
  \typeout{** the bibliography and then re-run LaTeX}
  \typeout{** twice to fix the references!}
  \typeout{******************************************}
  \typeout{}
 }


\end{document}